\documentclass[12pt]{iopart}

\expandafter\let\csname equation*\endcsname\relax
\expandafter\let\csname endequation*\endcsname\relax

\usepackage{bm}
\usepackage{csquotes}
\usepackage{mathtools}
\usepackage{xfrac}
\usepackage[dvipsnames]{xcolor}
\usepackage{graphicx}
\usepackage[ISO]{diffcoeff}
\usepackage{siunitx}
\usepackage{subcaption}
\usepackage{booktabs}
\usepackage{xspace}

\usepackage[colorlinks]{hyperref}
\hypersetup{%
	plainpages=true,
	breaklinks=true,
	hypertexnames=false,
	pageanchor=true,
	colorlinks=true,
	linkcolor={blue},
	citecolor={red},
	urlcolor={blue},
	anchorcolor={black}
}

\DeclarePairedDelimiter{\ket}{\lvert}{\rangle}
\DeclarePairedDelimiter{\aqty}{<}{>}
\DeclarePairedDelimiter{\pqty}{(}{)}
\DeclarePairedDelimiter{\bqty}{[}{]}
\DeclarePairedDelimiter{\vqty}{\vert}{\vert}
\DeclarePairedDelimiter{\Bqty}{\lbrace}{\rbrace}
\DeclarePairedDelimiterX{\braket}[2]{\langle}{\rangle}{#1\,\delimsize\vert\,\mathopen{}#2}
\DeclarePairedDelimiterX{\mel}[3]{\langle}{\rangle}{#1\,\delimsize\vert\,\mathopen{}#2\,\delimsize\vert\,\mathopen{}#3}

\DeclareMathOperator{\MAB}{MAB}
\DeclareMathOperator{\SAB}{SAB}
\DeclareMathOperator{\ISAB}{ISAB}
\DeclareMathOperator{\PMA}{PMA}
\DeclareMathOperator{\FF}{FF}
\DeclareMathOperator{\Sym}{Sym}
\newcommand{\eg}{\textit{e.g.}\xspace}
\newcommand{\ie}{\textit{i.e.}\xspace}

\newcommand{\supar}[1]{\ensuremath{^{\pqty*{#1}}}}
\newcommand{\software}[1]{\textsc{#1}}

\begin{document}

\title[Accelerating QMC with Set-Equivariant Architectures and Transfer Learning]{Accelerating Quantum Monte Carlo Calculations with Set-Equivariant Architectures and Transfer Learning}
\author{Manuel Gallego$^{1,2}$, Sebastián Roca-Jerat$^{1,3}$, David Zueco$^{1,3}$, Jes\'us Carrete$^{1,3}$}
\ead{jcarrete@unizar.es}
\address{$^1$ Instituto de Nanociencia y Materiales de Aragón (INMA), CSIC-Universidad de Zaragoza, Zaragoza 50009, Spain}
\address{$^2$ Departamento de Física Teórica, Universidad de Zaragoza, Zaragoza 50009, Spain}
\address{$^3$ Departamento de Física de la Materia Condensada, Universidad de Zaragoza, Zaragoza 50009, Spain}

\begin{abstract}
Machine-learning (ML) ansätze have greatly expanded the accuracy and reach of variational quantum Monte Carlo (QMC) calculations, in particular when exploring the manifold quantum phenomena exhibited by spin systems. However, the scalability of QMC is still compromised by several other bottlenecks, and specifically those related to the actual evaluation of observables based on random deviates that lies at the core of the approach. Here we show how the set-transformer architecture can be used to dramatically accelerate or even bypass that step, especially for time-consuming operators such as powers of the magnetization. We illustrate the procedure with a range of examples structured around quantum spin systems with long-range interactions, and comprising both regressions (to predict observables) and classifications (to detect phase transitions). Moreover, we show how transfer learning can be leveraged to reduce the training cost by reusing knowledge from different systems and smaller system sizes.
\end{abstract}

\vspace{2pc}

\submitto{Mach. Learn.: Sci. Technol.}

\maketitle

\section{Introduction}

The explosion of machine learning (ML) in the last two decades has had an extensive and deep impact on condensed matter physics and materials science, with every new advance in architectures and training quickly finding applications in those disciplines \cite{Carleo2019, Malica2025}. A good illustration is provided by the introduction of advanced regression models as interatomic potentials with ab-initio quality, starting with basic multilayer perceptrons \cite{behler} and moving towards sophisticated equivariant graph neural networks \cite{nequip}. Classification models have been equally prolific, with abundant examples of applications for the analysis of spectroscopic data, as a close analogue of image classification tasks \cite{classifier_examples}. Likewise, generative models have found many natural uses, some closely related to more mainstream tasks ---like the generation of candidate crystalline structures \cite{generative}, analogous to image or video generation--- and some field-specific ---such the creation of Boltzmann generators based on normalizing flows to efficiently sample the phase space of classical systems \cite{normalizing_flows}.

In the last years, ML strategies have also been deployed to assist in the QMC-based exploration of the rich physics of systems beyond the reach of analytical solutions. For instance, classifiers have been employed to detect phase transitions in problematic many-fermion systems \cite{Broecker2017, Chng2017} and in a more general setting \cite{ghosh2025learningphasesquantummonte}. Moreover, very general functional forms for regression and methods to train them efficiently have appeared on the scene \cite{Viteritti2023}, which can be used as flexible variational ansätze for a ground-state search through parametric minimization. The process is similar to ---and can therefore reuse part of the machinery developed for--- the stochastic minimization of a loss function when training a typical ML regression model, albeit with some important modifications. Firstly, the geometry of the quantum ground state problem favors stochastic reconfiguration \cite{PhysRevLett.80.4558,PhysRevB.64.024512}, a minimization method harnessing more information about the target function than standard ML minimizers like ADAM \cite{ADAM}. More importantly, the quantity being minimized is not a metric of the difference with the ground truth for a predefined set of inputs, but a physical quantity. Hence, the concerns about data availability that might be present in an ML regression setting are replaced by the complexity of sampling from the ground-state probability density function and evaluating the expected value of the Hamiltonian and other operators based on those samples. That complexity remains one of the causes of bottlenecks in variational QMC.

More specifically, both during and after training of a model for a finite-dimensional quantum system, the expected value of an operator $\bm{\mathcal{O}}$ in a state $\ket*{\psi}$ is evaluated using its expansion on the computational basis $\mathcal{C} = \Bqty*{\ket*{\nu}}$:

\begin{align}
\aqty*{\bm{\mathcal{O}}}_\psi &= \mel*{\psi}{\bm{\mathcal{O}}}{\psi} = \sum\limits_{\nu, \mu\in\mathcal{C}} \braket*{\psi}{\nu} \mel{\nu}{\bm{\mathcal{O}}}{\mu} \braket*{\mu}{\psi} = \nonumber\\
 &= \sum\limits_{\nu\in\mathcal{C}} \vqty*{\psi\pqty*{\nu}}^2\bqty*{\frac{1}{\psi\pqty*{\nu}}\sum\limits_{\mu\in\mathcal{C}}\psi\pqty*{\mu}\mel{\nu}{\bm{\mathcal{O}}}{\mu}} \,\label{eqn:operator}
\end{align}
\noindent with $\psi\pqty*{\mu} = \braket*{\mu}{\psi}$.
In a QMC context, $\aqty*{\bm{\mathcal{O}}}_\psi$ is estimated from a finite sample of $\ket*{\nu}$ randomly extracted from $\mathcal{C}$ with a Markov chain that approximates the distribution induced by $\psi$. The term in square brackets is known as the local estimator $\mathcal{O}_\nu$, and in practice only involves a sum over the small subset of $\mu\in\mathcal{C}$ connected to each $\nu$ through $\bm{\mathcal{O}}$. For operators built as linear combinations of Pauli strings \cite{PauliStrings} the relevant $\mu$ are easy to enumerate by exploring all possible spin orientations at the sites affected by each string. For instance, when evaluating the local estimators of the total magnetization along OZ (the quantization axis), $\bm{\mathcal{M}\supar{z}}=\sum\limits_{i=1}^N\bm{\sigma}_i\supar{z}$, for a system of $N$ spin-$\sfrac{1}{2}$ sites, only $\ket*{\mu} = \ket*{\nu}$ makes a nonzero contribution, whereas to evaluate $\bm{\mathcal{M}\supar{x}}=\sum\limits_{i=1}^N\bm{\sigma}_i\supar{x}$, the $N$ states obtained from $\ket*{\nu}$ by a single spin flip have to be taken into account. Therefore, evaluating the expectation value of operators involving more sites incurs a geometrically increasing cost, with important implications for the extensibility of the method. As an example, if $\bm{\mathcal{H}}$ is the Hamiltonian of a $J_1$-$J_2$ Heisenberg model, then each $\mathcal{H}_\nu$ involves $N^2$ states, and thus obtaining the expectation value of $\bm{\mathcal{H}}^3$ required to refine an estimate of the ground state through a single Lanczos iteration has $N^6$ complexity, rendering the practicality of the procedure questionable and prompting the design of workarounds \cite{Chen2024}. This is compounded by the fact that, in short-ranged models, successive powers of $\bm{\mathcal{H}}$ are progressively less sparse, increasing the cost of each individual matrix element as well.

In this paper we present an accelerated approach to the calculation of expectation values by making more efficient use of the samples drawn from $\vqty*{\psi}^2$. A critical feature of such approaches is that the estimate of an expectation value cannot depend on the order in which the $\ket*{\nu}$ are drawn from $\vqty*{\psi}^2$ or fed to the model, \ie, it must be permutation invariant. Moreover, they must be able to work on collections of $\ket*{\nu}$ with different cardinality. Those are the characteristics of a set-invariant architecture. We have therefore based our strategy on an architecture developed as a universal, efficient approximation to set-equivariant functions, the set transformer \cite{set_transformer}. We use it as the backbone of permutation-invariant regression models. We also show how to build a classifier based on the set transformer to detect phase transitions. The calculation of each expectation value using the training architecture is three to four orders of magnitude faster than the direct approach. In addition to analyzing the performance of these models, we discuss how to build them more efficiently by amortizing the training cost over ranges of Hamiltonian parameters, numbers of samples, and system sizes: for instance, after pretraining for a chain of $50$ spins, extending the model to $100$ and $150$ spins only requires an additional $70\%$ computational effort.

The next section gives the details of our methods. We then present the results from our regression and classification experiments; finally, we discuss the conclusions.

\section{Method}\label{sec:methodology}

\subsection{The Set-Transformer Architecture}

Our goal is to build regression models $f_{\mathcal{O}}\pqty*{\mathcal{S}\subset\mathcal{C}}$ to efficiently evaluate MC expectation values of the form expressed by~\eqref{eqn:operator} over a random sample $\mathcal{S}$ from the computational basis $\mathcal{C}$. We also aim to develop classification models taking $\mathcal{S}$ as their input and emitting a verdict about the phase of the system (\ie, to answer whether it is ferromagnetic or not). A key design consideration of such models is that they treat $\mathcal{S}$ as a set, \ie, the answer can only depend on which elements of $\mathcal{C}$ are included in $\mathcal{S}$. Therefore, the functional form of the regressions and classifications must allow for a variable number of samples in $\mathcal{S}$ and be completely permutation invariant, \ie, $f_{\mathcal{O}}\bqty*{\tau\pqty*{\mathcal{S}}} = f_{\mathcal{O}}\pqty*{\mathcal{S}}$ for any permutation $\tau\in\Sym\pqty*{\mathcal{S}}$.

The deep-sets theorem \cite{deep_sets} states that a universal form of a permutation-invariant $f_{\mathcal{O}}\pqty*{\mathcal{S}}$ is

\begin{equation}
    f_{\mathcal{O}}\pqty*{\mathcal{S}} = G\bqty*{\sum\limits_{\nu\in\mathcal{S}}g\pqty*{\ket{\nu}}}.
\end{equation}

\noindent In principle, a permutation-invariant regression model could hence be built by choosing and training sufficiently general ML models to take the roles of $g$ and $G$. When dealing with sampled spin configurations $\ket*{\nu}\in\mathcal{C}$, these deep-sets (or set-pooling) architectures take the form of conventional (\eg fully connected or convolutional) models that are applied to each $\ket*{\nu}$ independently, and whose results are summed or averaged to produce the final output. This scheme has also been successfully exploited in ML force fields \cite{mlff_uncertainties}, where the inputs are atomic coordinates and types, the output is the system's potential energy, and $G$ is taken as the identity to create an additive architecture. However, in a more general setting the scalability and expressive power of actual realizations of this construction based on finite models have been shown to be limited \cite{deep_sets_limitations}, and the same applies to slight variations using set-pooling functions other than the sum.

A more practically oriented approach to permutation invariance that still results in an universal approximator of set-invariant functions is the set transformer \cite{set_transformer}, derived from the extremely successful transformer architecture \cite{attention}. The set transformer can be described starting from a conventional multihead dot-product attention block with LayerNorm normalization \cite{LayerNorm} but without any positional encoding \cite{Shaw2018} or dropout \cite{dropout}. We will denote that block as $\MAB\pqty*{\bm{X}, \bm{Y}}$, where $\bm{X}$ and $\bm{Y}$ are two arrays of $n$ $d$-dimensional tokens each, with $\bm{X}$ being used to generate the queries and $\bm{Y}$ being used to generate the keys and values of the attention mechanism. This basic $\MAB$ block is then used to build two different kinds of higher-level components: set-attention blocks ($\SAB$) and pooling-by-multihead-attention ($\PMA_k$) blocks. While $\PMA_k$ blocks are set invariant, $\SAB$s are set equivariant, \ie, $\SAB\bqty*{\tau\pqty*{\bm{X}}} = \tau\bqty*{\SAB\bqty*{\bm{X}}}\;\forall \tau\in\Sym\pqty*{\bm{X}}$. A full set transformer contains an encoder pipeline in the form of a stack of set-equivariant blocks followed by a decoder stage producing the final output. In our models we design the decoder stage to be set invariant, thus making the whole transformer invariant as well.

A set-equivariant block can be straightforwardly built by turning the multihead attention block into a self-attention block by short-circuiting its inputs [$\SAB\pqty*{\bm{X}} = \MAB\pqty*{\bm{X}, \bm{X}}$]; in fact, it is this well-known property of the dot-product self attention that motivates the introduction of positional encodings in applications where neighborhood relations between input tokens are important and therefore set equivariance is undesirable. However, the scalability of this simple $\SAB$ is limited and a more performance-oriented alternative is the induced set-attention block $\ISAB_m\pqty*{\bm{X}}=\MAB\bqty*{\bm{X}, \MAB\pqty*{\bm{I}, \bm{X}}}$, where $\bm{I}$ is a set of $m < n$ trainable inducing points.

Regarding the invariant decoder stage, its key component is a block that performs pooling by multihead attention, specified as $\PMA_k\pqty*{\bm{X}} = \MAB\bqty*{\bm{S}, \FF\pqty*{\bm{X}}}$, where $\bm{S}$ is an array of $k$ trainable seed vectors and $\FF$ is an arbitrary feed-forward block acting row by row. Further processing can be done on the output of this block, for instance by applying an $\SAB$.

\begin{figure}
    \begin{center}
    \begin{tabular}{c}
    \multicolumn{1}{l}{Multihead attention block (MAB):}\\
    \includegraphics[scale=.5]{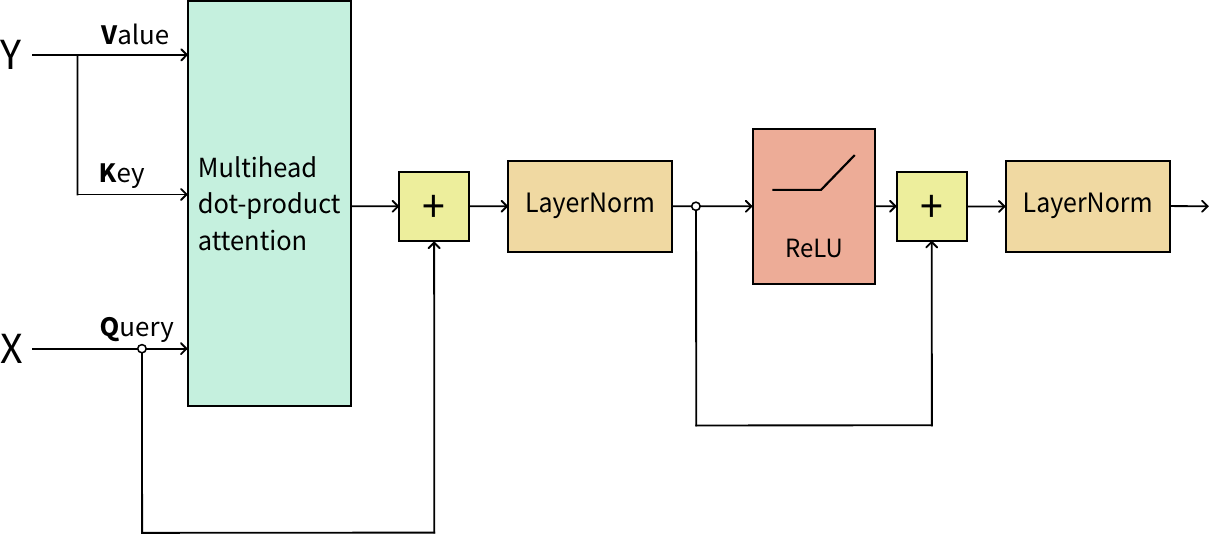}\\
    \multicolumn{1}{l}{Induced set attention block (ISAB):}\\
    \includegraphics[scale=.5]{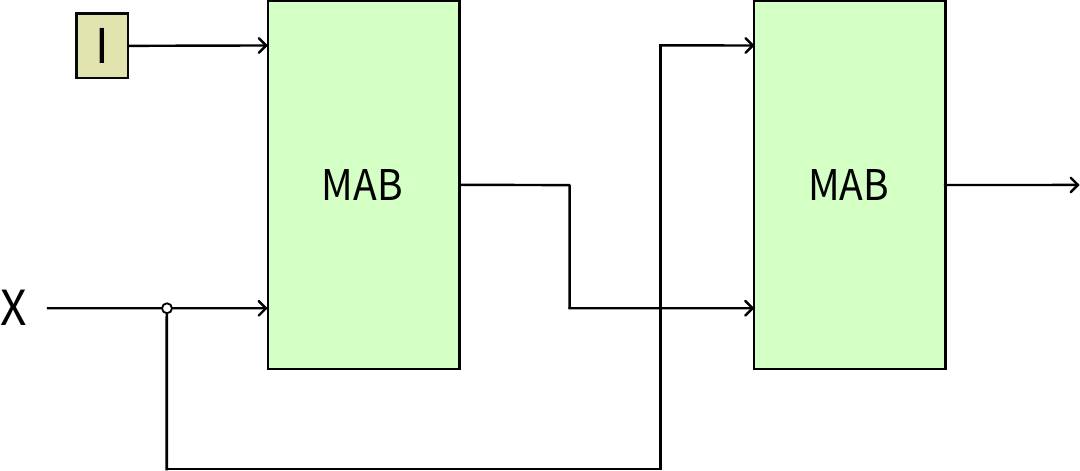}\\
    \multicolumn{1}{l}{Pooling by multihead attention (PMA):}\\
    \includegraphics[scale=.5]{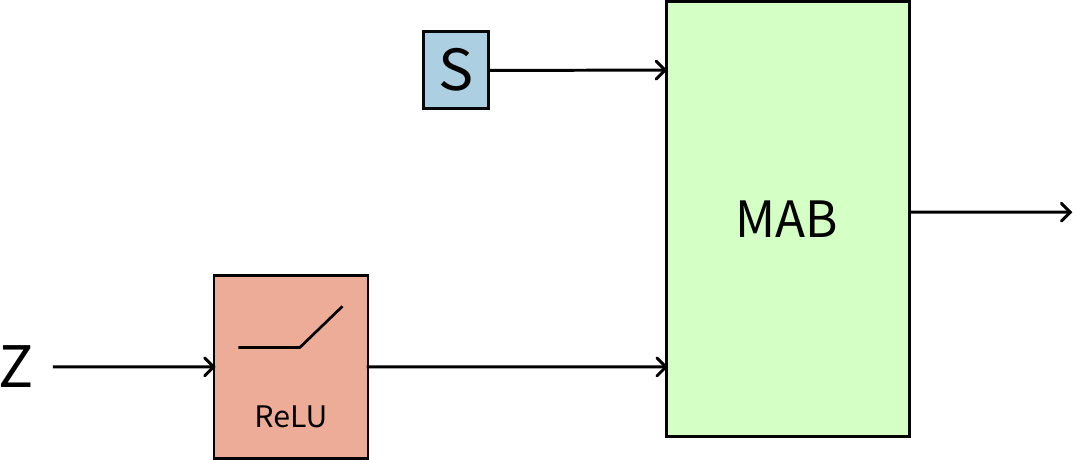}
    \end{tabular}
    \end{center}
    \caption{Structure of the main building blocks of the set transformer. The multihead attention block (MAB) is the core component, defined as a standard multihead dot-product attention with LayerNorm and ReLU, but without positional encoding or dropout. The induced set-attention Block (ISAB) is an alternative to self-attention that preserves the set equivariance but improves scaling; it is constructed as $\ISAB_m\pqty*{\bm{X}} = \MAB\bqty*{\bm{X}, \MAB\pqty*{\bm{I}, \bm{X}}}$ with $m < n$ trainable inducing points $\bm{I}$. The pooling by multihead attention (PMA) block defines the set-invariant decoder stage, specified as $\PMA_k\pqty*{\bm{X}} = \MAB\bqty*{\bm{S}, \FF\pqty*{\bm{X}}}$, where $\bm{S}$ are $k$ trainable seed vectors and $FF$ is a feed-forward block.}
    \label{fig:set_transformer_blocks}
\end{figure}

For the system sizes employed in this work, we do not need to resort to the ISAB. Our set transformers are built entirely out of self-attention and pooling-by-self-attention blocks. The whole set of building blocks is schematically represented in Figure~\ref{fig:set_transformer_blocks}; following the original recipe \cite{set_transformer}, in this work we employ implementations of the set transformer where each processing component is simply a non-linear unit, more specifically a ReLU \cite{ReLU}.

It is important to note that a model whose inputs are sampled configurations of spins disregards all information about the phase of the wave function and, more specifically, cannot discriminate between two wave functions $\ket*{\psi}$ and $\ket*{\phi}$ such that $\vqty*{\psi\pqty*{\mu}} = \vqty*{\phi\pqty*{\mu}}\,\;\forall\mu\in\mathcal{C}$. For general complex wave functions the missing information can be incorporated into the input by appending the full complex value of $\psi\pqty*{\mu}$ after the spin configuration; however, in this work, where our goal is to introduce and validate the method, we focus on Hamiltonians for which this extension is not necessary. This is partly because the relevant pieces of information can be learned during training, and partly because of the connection between modulus and sign changes for real wave functions that behave smoothly with respect to a parameter of the Hamiltonian.

A set-transformer-based approach is model-agnostic, and can be applied in other kinds of QMC. The only requirement is access to configurations drawn from a sampling measure, be they fermionic samples, occupation snapshots, Hubbard–Stratonovich slices, or equal-time Green matrices. For instance, in Ref.~\cite{Broecker2017}, Broecker et al. demonstrated that a CNN trained on equal-time Green matrices produced by auxiliary-field QMC can locate  fermionic phase transitions even when conventional sign-reweighted estimators fail. Their key idea is to approximate the \enquote{state function} directly from QMC configurations, thereby bypassing explicit sign reweighting in the final prediction. This complements our set-equivariant approach: a set transformer can be trained with exactly the same sampled objects and perform phase classification without computing observables, thus inheriting the same practical insensitivity to the sign problem for classification tasks.

\subsection{Our Regression and Classification Models}\label{subsec:architectures}

We use two specific set-transformer models, one for regression and one for classification. Our classifier contains an encoder stage made up of a $5$-head $\SAB$ followed by a $10$-head SAB, a decoder stage comprising a $5$-head, single-seed $\PMA$ followed by two $10$-head $\SAB$s with a LayerNorm between them, a linear layer to generate the logits, and a SoftMax layer to generate the final output. We do not include any $\ISAB$ block since the quadratic scaling of attention does not become a problem for these system sizes. When we apply this model to label phases as antiferromagnetic, ferromagnetic or paramagnetic, we obtain better results by using a sequence of two binary classifiers than by resorting to a direct ternary classifier. The first of those classifiers sets aside paramagnetic states; the remaining ones are passed on to the second classifier, which then discriminates between ferromagnetic and antiferromagnetic states.

The inputs to the classifier are arrays with dimensions $n_{\mathrm{samples}}\times N$ and values $\pm 1$ representing the two possible spin orientations. Each sample is divided into fixed-sized tokens that make up the basic units the NNs act on. An attention mask is used across all models to allow inputs with different numbers of samples up to a maximum. The remaining input slots are filled with padding values and are guaranteed not to be attended to.

Our regression model is similar, albeit with two differences: first, the SoftMax layer is absent and the linear layer preceding it has a single output; second, a trainable embedding layer of width $2$ is inserted immediately after the input. This embedding improves the performance of the model in the context of transfer learning where, as detailed later, the computational basis is augmented with a third possible value for each individual spin meaning \enquote{not present}.

\subsection{Hamiltonian, Ansatz and Ground-State Search}

Our model system is a one-dimensional chain of $N$ spins with $s=\sfrac{1}{2}$, periodic boundary conditions and long-range interactions expressed by the Hamiltonian

\begin{equation}
\bm{\mathcal{H}} = \sum\limits_{ij}^N J_{ij} \bm{\sigma}_i\supar{z} \bm{\sigma}_j\supar{z}\ - h \sum_i^N \bm{\sigma}_i\supar{x}.
\label{eqn:ising:H}
\end{equation}

\noindent Here, $h$ is an external magnetic field and the coupling between spins decays following a power law

\begin{equation}
J_{ij} = \begin{cases}
    \frac{J}{\eta} & \text { if } \; i=j \\
    \frac{J}{\eta\vqty*{i-j}^{\alpha}} & \text { otherwise},
    \end{cases}
\label{eqn:ising:J}
\end{equation}

\noindent where $\vqty*{i-j}$ must be understood in the context of the periodic boundary conditions and $\eta=1+\sum\limits_{j=1}^Nj^{-\alpha}$. Thus, the three key parameters of the model are
the size, \ie the number os spins, $N$; the ratio $J/h$ that controls the strength of the interactions between spins, and $\alpha$, which determines how fast those interactions decay with distance. The role of the Kac renormalization factor, $\eta$, is to render the model extensive  with respect to $N$ \cite{campa2009statistical, romanroche2023} .

We choose this model because it provides a tunable interpolation between two paradigmatic and exactly solvable limits: the transverse-field Ising model with nearest-neighbor interactions, recovered in the limit $\alpha \to \infty$, and the fully-connected Lipkin-Meshkov-Glick (LMG) model at $\alpha = 0$. The former is not only analytically tractable but also serves as the standard benchmark for quantum simulations using both machine learning and other numerical techniques. The latter is also exactly solvable and well understood, offering a complementary point of comparison. By continuously varying $\alpha$, the model exhibits a rich phenomenology arising from the interplay between interaction range and quantum fluctuations. For $\alpha > \alpha^*$, the system behaves effectively as short-range, while for $\alpha \in (0,1)$ strong long-range effects emerge, with $\eta$ diverging as a function of system size $N$, thus becoming essential to recover the thermodynamic limit as $N \to \infty$. In between those, a weak long-range regime appears. Moreover, the model has recently been studied across the full range of $\alpha$ using a variety of methods including QNNs \cite{koziol2021quantumcritical, Kim2024, PhysRevB.110.205147, HerraizLpez2025}, enabling us to benchmark our results and quantify the numerical speedup achieved by our approach.

To find the ground state of the Hamiltonian in~\eqref{eqn:ising:H}, we optimize the parameters of a neural-quantum-state ansatz. In particular, we choose a flexible functional form based on the vision transformer (ViT \cite{vision_transformer, Viteritti2023}), another widely used specialization of the transformer architecture. In our model, the projection of state $\ket*{\psi}$, represented by the ansatz, on element $\ket{s}$ of the computational basis is determined through the following steps:

\begin{enumerate}
    \item $\ket{s}$ is encoded as a length-$N$ floating-point array $\mathbf{x}_{\mathrm{in}}$ with values $\pm 1$.
    \item $\mathbf{x}_{\text{in}}$ is split into a sequence of length-$d$ tokens.
    \item Each token is processed by a trainable, linear embedding layer into a length-$d_{\text{emb}}$ array.
    \item The processed tokens are normalized by a LayerNorm \cite{LayerNorm} and fed into an $\MAB$ with $H$ heads. The attention models the interaction between tokens. To make this interaction respect translational symmetry, the attention is parameterized by a circulant matrix.
    \item The results are further processed, in parallel, by a multilayer perceptron (MLP).
    \item The previous two steps are repeated a preset number of times. Each independently trainable combination of a $\MAB$ and an MLP is termed a \enquote{core block} of the model
    \item The results of the last core block are passed through a simple pooling operation consisting of a log-cosh layer and a sum over tokens.
    \item That unified output is processed by a last MLP and a final layer that multiplies the result by a trainable factor and adds a trainable offset to generate an unnormalized version of $\braket*{s}{\psi}$.
\end{enumerate}

As in the case of the set transformer, within each core block, two ResNet-style shortcuts \cite{ResNet} allow the network to be trained to bypass the mixing and processing steps if necessary to improve performance and therefore avoid the drawbacks of increased network depth. To enforce translation symmetry also at the intra-token level, the sequence above is repeated $d$ times with the input shifted by one position each time with respect to the previous one, so that every spin occupies all possible positions within each token, and the final output is taken as the average of those $d$ runs.

The architecture employed to find the ground states is defined by the set of parameters $d = N/10$, $d_{\text{emb}} = 14$ and $H = 2$. The MLP described in point (v) of the list above is made up of $3$ layers. Each of them consists of a dense layer to which a LayerNorm operation is applied, followed by the Swish activation function \cite{swish}. Finally, the last MLP ---from point (viii)--- is a single layer, like the ones just described, that reduces the dimension from $d_{\text{emb}} = 14$ to $d_{\text{MLP}}=5$. This neural-quantum-state ansatz is constructed using the \software{Flax} neural-network library \cite{flax}, running on top of \software{JAX} \cite{jax2018github}. The training is carried out through the library \software{NetKet} \cite{netket3:2022}, which also takes care of the MC sampling. A total of $500$ iterations are used to find the ground states. In each of these iterations, the expected value of $\bm{\mathcal{H}}$ is estimated according to~\eqref{eqn:operator} using $4096$ samples spread over $1024$ independent Markov chains. The optimization protocol chosen is stochastic gradient descent (SGD) with a learning rate $\lambda$, combined with the stochastic reconfiguration (SR) method \cite{PhysRevB.61.2599}, characterized by a stabilizing parameter called diagonal shift that we denote as $\Delta_{\mathrm{sr}}$. For both $\lambda$ and $\Delta_{\mathrm{sr}}$ we define custom schedules with \software{Optax} \cite{deepmind2020jax}. The protocol designed for $\lambda$ is defined by an initial value of the learning rate, $\lambda_0 = 0.1$, a maximum value $\lambda_{\mathrm{max}} = 5.0$ that it attains after $n_{\mathrm{warm}} = 75$ iterations with a linear warm-up, and an exponential decay with ratio $\gamma = 0.995$. For the diagonal shift $\Delta_{\mathrm{sr}}$ we employ a linear schedule with an initial value of $\Delta_{\mathrm{sr}} = \num{e-2}$ and a final value of $\Delta_{\mathrm{sr}} = \num{e-4}$. The code that reproduces the ground state generation is available on GitHub \cite{vit_github}.

\subsection{Data Generation and Training Procedures}\label{sec:datagentrain}

In order to train the set transformers for the different tasks defined above, we generate ground states from different configurations of \eqref{eqn:ising:H} following the procedure outlined in the previous section. 

In the case of the classification task, we run calculations for the whole gamut of model parameters for chains of $50$ spins, covering values of $\alpha$ from $0$ to $6$ with a step size $\delta\alpha = 0.4$, thus creating a reference for the ground truth of the model in all its complexity. We also explore values of $J/h$ from $-5$ to $10$, with $\delta \pqty*{J/h} = 0.375$, which is likewise enough to sample all the magnetic phases of the system.

To perform the training of the classifier architecture, we use only states with $\alpha \in \Bqty*{0, 3.2, 6}$. We augment the original training set, $T$, increasing the number of its elements by a factor of five, through the following subsampling aggregation scheme:

\begin{enumerate}
    \item At each iteration, we draw a random element $t\in T$, and a uniform random integer $c\in\bqty*{1, 1024}$.
    \item We extract a random subsample of $c$ elements from the input part of $t$.
    \item We add that subsample, along with the output part of $t$, to the new training set.
\end{enumerate}

This procedure not only enlarges, but also enriches the set with elements of different cardinality, whose correct treatment is one of the key features we are trying to promote in our models. This same technique is applied to the validation and test sets. The learning process is driven by an Adam optimizer with a fixed value of $5\times 10^{-5}$ for the learning rate and had a total duration of $500$ epochs.

Regarding the regression targets, we define the generalized magnetization as
\begin{equation*}
\bm{\hat \bm{m_z}} = \begin{cases}
\frac{1}{N} \sum_{i=1}^N \sigma_{z,i} & \quad J<0, \\
\frac{1}{N} \sum_{i=1}^N \left( -1 \right)^i\sigma_{z,i} & \quad J>0.
\end{cases}
\end{equation*} 

\noindent The two cases correspond to the regular and staggered magnetizations, respectively.  We train three networks to predict the first and second moments of this quantity, $\aqty*{\bm{\hat m_z}^2}$ and $\aqty*{\bm{\hat m_z}^4}$, as well as the second Rényi entanglement entropy $\bm{\hat S_2} = -\log_2\Bqty*{\bm{\rho_A}^2}$. All three networks have the same architecture, described in (\ref{subsec:architectures}). We begin by analyzing the second and fourth moments of the generalized magnetization over a broad range of values for $\alpha$ and $J$. The focus of this analysis is on the overall behavior of these quantities, rather than on the detailed features near phase transitions, which are addressed separately. The models are trained using system states corresponding to the same set of $\alpha$ values used for the classifier. To ensure that the network remains effective across configurations with varying total sample counts, data augmentation is applied to the training, validation, and test sets.

\begin{figure}[t]
    \centering
    \includegraphics[width=0.8\linewidth]{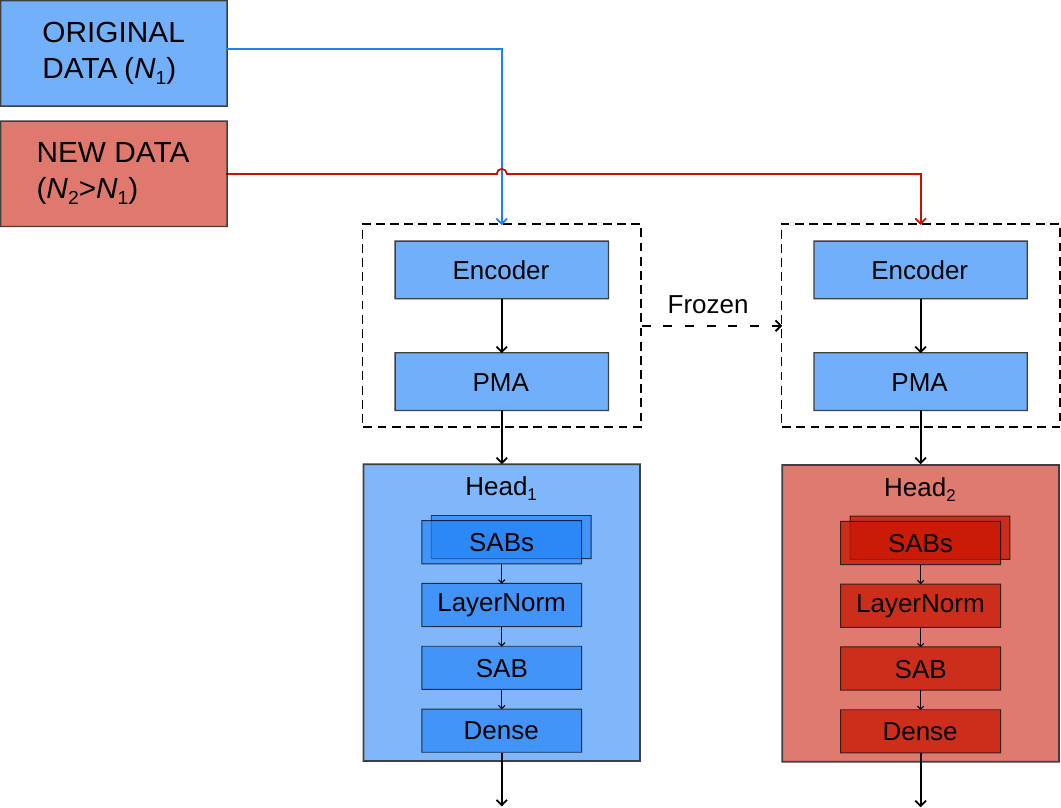}
    \caption{Workflow of the transfer learning scheme. The model is first trained on data from the smaller system. A new model then inherits a subset of the trained parameters, which are frozen to prevent further updates. The remaining parameters are subsequently trained on data from the larger system. \enquote{Dense} refers to a fully connected block, whose precise form will be different for regressions and classifications.}
    \label{fig:transfer-learning}
\end{figure}

\begin{figure}[t]
    \centering
    \includegraphics[width=0.8\linewidth]{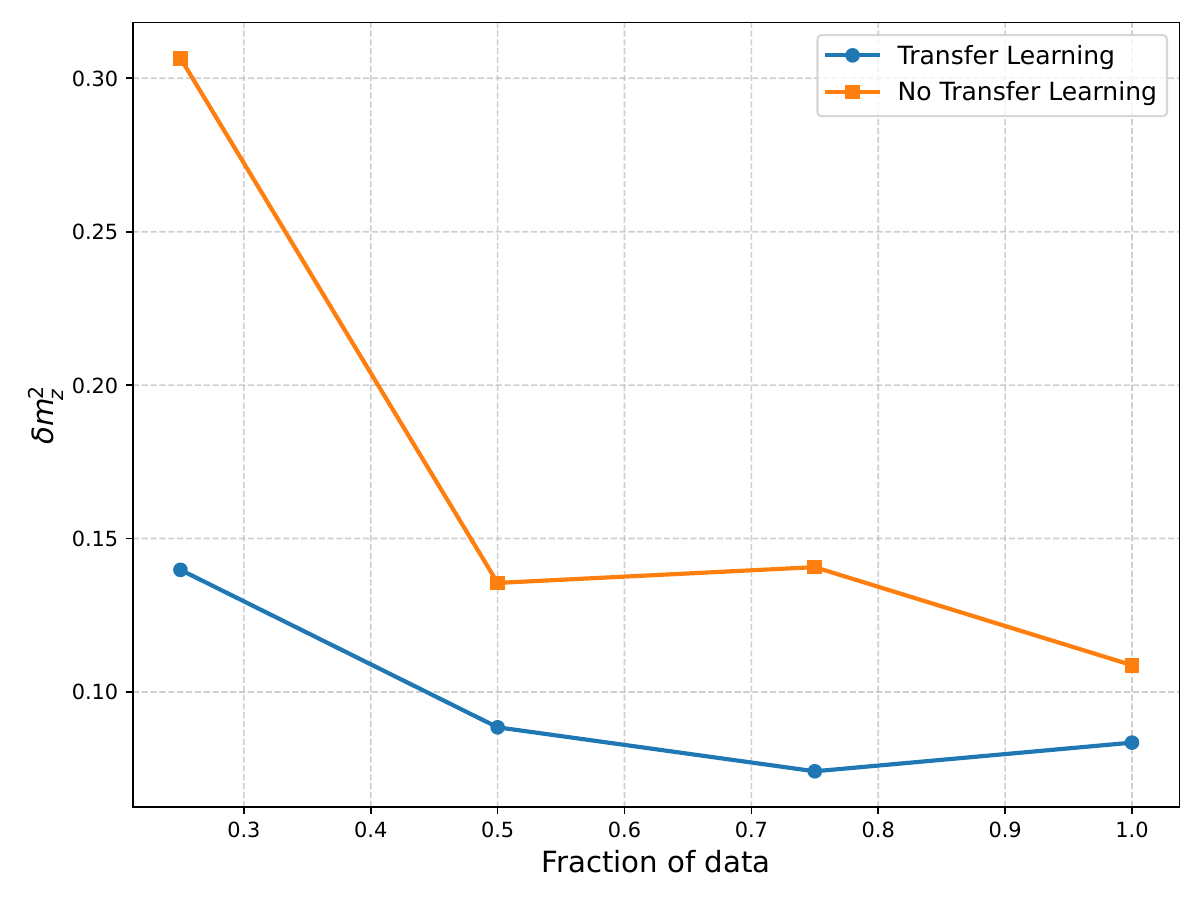}
    \caption{Accuracy of the predictions with and without transfer learning for $N=100$ and $\alpha=2.5$. The standard error over the validation set is computed as $\delta m^2_z = \sqrt{\frac{1}{n-1}\sum_{i=1}^n\pqty*{m_{z,i}^2-\tilde{m}_{z,i}^2}^2}$, where $n$ is the number of data points in that set, $m_{z,i}^2$ are the fluctuations of the magnetization calculated via QMC with original variational states and $2048$ samples, and $\tilde{m}_{z,i}^2$ are the set-transformer predictions for the same quantities.}
    \label{fig:acc_data_fraction}
\end{figure}

The second stage in the application of the regressor model consists in analyzing the behavior of the expected values of both the second moment of the magnetization and the entanglement entropy in the vicinity of the critical point. To this end, we generate new reference ground states near the phase transition for each system with $\alpha \in \lbrace 1.66, 1.7, 2.5, 6 \rbrace$, considering spin chains of lengths $50$, $100$, and $150$~\cite{PhysRevB.110.205147}. This finer grid allows us to capture the intricate features of the ground state and the order parameter across the transition, which constitute the most challenging regions of the parameter space. For training, we only use the states corresponding to the extreme values $\alpha \in \lbrace 1.66, 6 \rbrace$ and apply data augmentation to enrich the dataset. For both observables, optimization was performed using stochastic gradient descent with a polynomial learning rate schedule.

The problem of calculating an accurate value for entanglement entropies, such as the second Rényi entanglement entropy in our case, remains challenging even within the QMC toolbox, since results usually present large fluctuations which may obscure properties like scaling laws with system size. A common workaround is to perform several runs for each case and extract more robust statistics. Here, we propose a similar procedure: for the training process, we perform the full QMC calculation five times for each pair $\lbrace J/h, \alpha\rbrace$ and obtain the corresponding five values of $\hat S_2$. Then, we train the set transformer with these five estimates of the ground state, but all are labeled with the mean of these five values of $\hat S_2$, instead of the individual calculations. The goal of this technique is to prevent the network from overfitting to unphysical fluctuations arising from QMC calculations.

\subsection{Transfer Learning}\label{sec:transferlearnign}

\begin{figure}[t]
    \centering
    \includegraphics[width=0.8\linewidth]{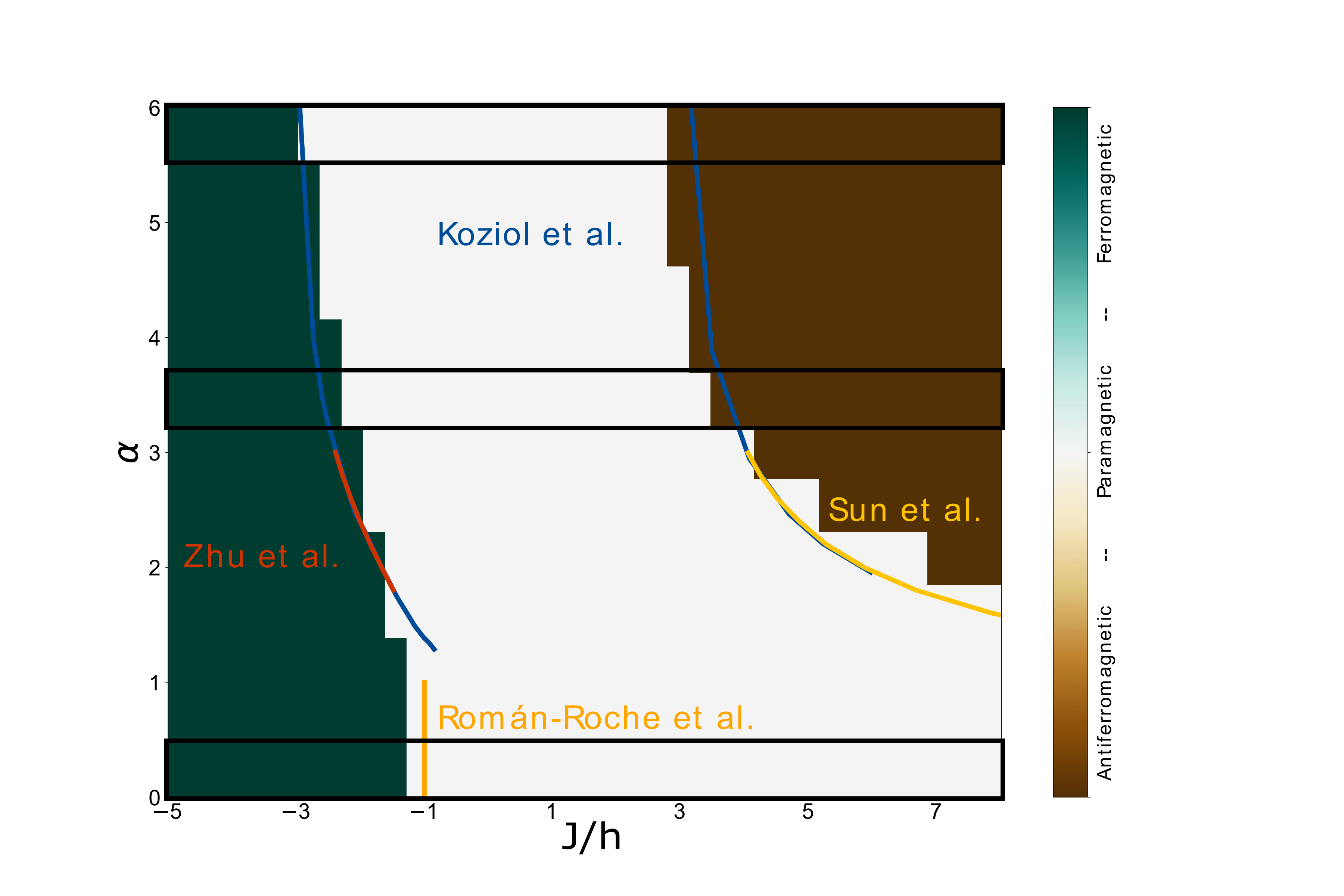}
    \caption{Results of the nested binary classification of states into paramagnetic, ferromagnetic and antiferromagnetic categories for a system of $50$ spins throughout the ranges of $\alpha$ and $J$. This model was trained on $\alpha\in\Bqty*{0, 3.2, 6}$, highlighted with black rectangles, with up to $1024$ samples per training point. Also shown are the phase boundaries extracted from Refs.~\cite{romanroche2023, zhu2018fidelity, koziol2021quantumcritical, sun2017}}
    \label{fig:classifier}
\end{figure}

Accelerated classification and regression methods are particularly advantageous in the case of operators connecting large subsets of states that make the evaluation of~\eqref{eqn:operator} particularly expensive. Generating data to train these models is itself an expensive operation, especially for high-dimensional systems. It is therefore desirable to be able to leverage the commonalities between models of different sizes in the same family, generating training data predominantly from the smaller members of that family.

To make that kind of workflow possible, we need the ML model $f_{\mathcal{O}}\pqty*{\mathcal{S}\subset\mathcal{C}}$ to accept sets $\mathcal{S}$ with samples of different length (although each $\mathcal{S}$ is still homogeneous). Note that this is a different and additional requirement than just being able to accept sets with different cardinals, which is accomplished using attention masks. To allow the model to work with different numbers of spins we allow three spin values in the input, $\Bqty*{0, 1, 2}$, with $0$ and $1$ playing the roles previously assigned to $\pm 1$ and $2$ representing missing spins. We then add a trainable embedding layer after the input to map those to more optimal values for the ML models. We do not use masking along this dimension, but let the networks learn how to interpret the ternary input instead. 

Figure \ref{fig:transfer-learning} illustrates the transfer learning workflow employed in this study. The network is first trained on data from the smaller system (50 spins). After this initial stage, the gradients of the main architectural block are frozen to prevent further updates, and only the final heads of the network are retrained from scratch using data from the larger systems (100 and 150 spins). Each of these subsequent training stages requires approximately one third less time than the initial training.

Although transfer learning does not enable the exploration of entirely new system sizes, as data from the target systems is still required, it improves efficiency by reducing the amount of data needed compared to training a network from scratch. Figure~ \ref{fig:acc_data_fraction} illustrates this fact through a direct comparison of the accuracies of networks trained to predict the fluctuations in the magnetization, with or without transfer learning, as a function of the fraction of the available data used in their training.

\section{Results}

\begin{figure}[t]
    \centering
    \includegraphics[width=0.95\linewidth]{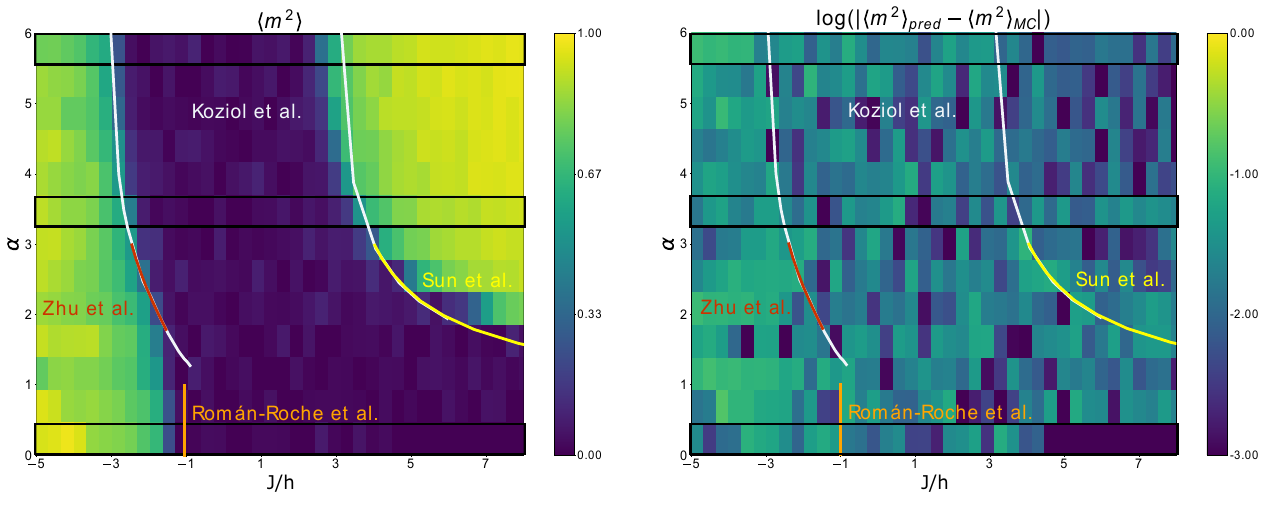}
    \caption{Regression results for the second moment of the magnetization for a system of $50$ spins throughout the ranges of $\alpha$ and $J$. \textbf{Left:} $\langle m^2 \rangle$ predicted by the set transformer. \textbf{Right:} difference between the predictions and the results obtained via QMC calculations, in logarithmic scale. This model was trained on $\alpha\in\Bqty*{0, 3.2, 6}$, highlighted with black rectangles, with up to $1024$ samples per training point. Also shown are the phase boundaries extracted from Refs. \cite{romanroche2023, zhu2018fidelity, koziol2021quantumcritical, sun2017}.}
    \label{fig:m2}
\end{figure}

We first present the results of the phase-detection classifier in Figure~\ref{fig:classifier}, which also depicts the phase boundaries proposed in Refs.~\cite{romanroche2023, zhu2018fidelity, koziol2021quantumcritical, sun2017}, color coded according to their source. Despite being trained on only three values of $\alpha$, the classifier works well across the whole spectrum of $\alpha$. The border between the ferro- and paramagnetic regions is particularly well defined, while a few artifacts are visible in the case of the threshold between para- and antiferromagnetic ground-state configurations. Those may be partially attributable to a problem in the training data. The erroneous predictions are well separated from the threshold itself and can easily be identified as such.

Closely related to this classification is the regression model for $\aqty*{m^2}$, whose results are shown in the left panel of Figure~\ref{fig:m2}.  The transitions among the three phases are again immediately visible and agree with previous results. Furthermore, the logarithm of the difference between the predicted results and those calculated with QMC is shown in the right side of Figure~\ref{fig:m2}. The model seems to perform worst in the ferromagnetic phase, where errors can reach $\sim 15\%$, than in the other two phases of the system, and are primary located within certain bands, \eg $\alpha\in\lbrace 1, 2, 6\rbrace$. This correlates with the worse quality of the training data in the ferromagnetic region. The training process takes approximately 15 minutes for the first system size and about 11 minutes for each subsequent one. Once training is complete, the evaluation of an individual expectation value requires only \SI{0.04}{\second}, which amounts to a speed-up of roughly $\times 100$ for $N=50$ and $\times 1800$ for $N=100$.

\begin{figure}[t]
    \centering
    \includegraphics[width=0.95\linewidth]{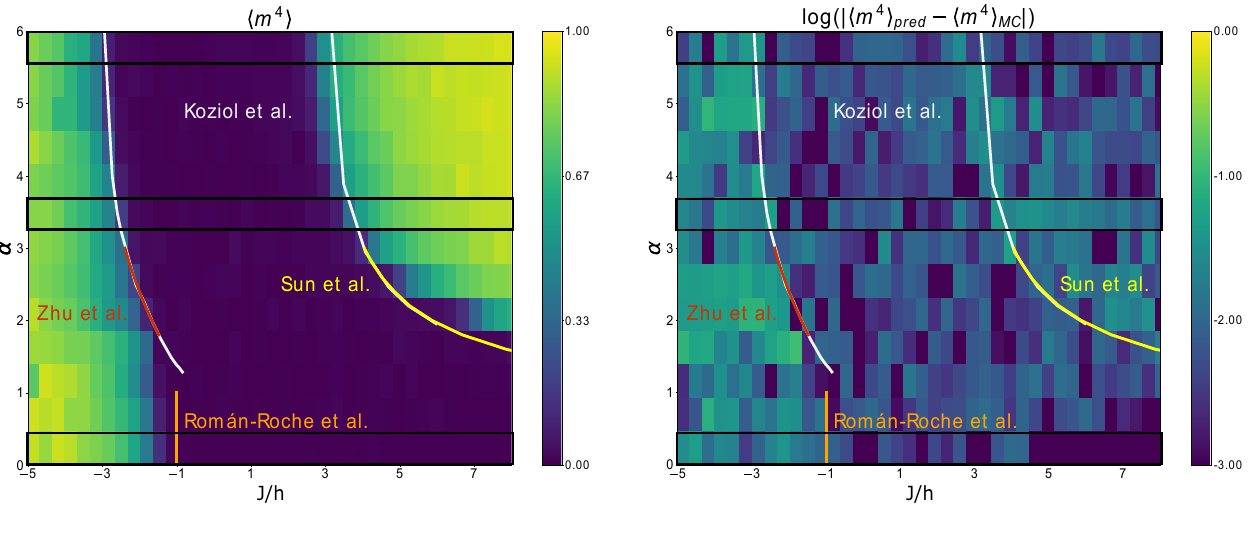}
    \caption{Regression results for the fourth moment of the magnetization for a system of $50$ spins throughout the ranges of $\alpha$ and $J$. \textbf{Left} $\langle m^4 \rangle$ predicted by the set transformer. \textbf{Right} difference in logarithmic scale between the predictions and the results obtained via QMC calculations. This model is trained on $\alpha\in\Bqty*{0, 3.2, 6}$, highlighted with black rectangles, with up to $1024$ samples per training point. Also shown are the phase boundaries extracted from Refs. \cite{romanroche2023, zhu2018fidelity, koziol2021quantumcritical, sun2017}.}
    \label{fig:m4}
\end{figure}

We now turn our attention to the fourth moment of the magnetization, a quantity that is computationally demanding to evaluate using QMC techniques for the reasons outlined in Section~\ref{sec:methodology}. As shown in the left panel of Figure~\ref{fig:m4}, the set-transformer predictions remain accurate, and the right side of Figure~\ref{fig:m4} displays the logarithmic error with respect to the QMC results. Importantly, despite the higher computational cost associated with this observable, we observe no degradation in predictive performance compared to the magnetization. Again, we can see the worst predictions appear in the ferromagnetic phase of the system.

\begin{figure}[t]
    \centering
    \includegraphics[width=0.8\linewidth]{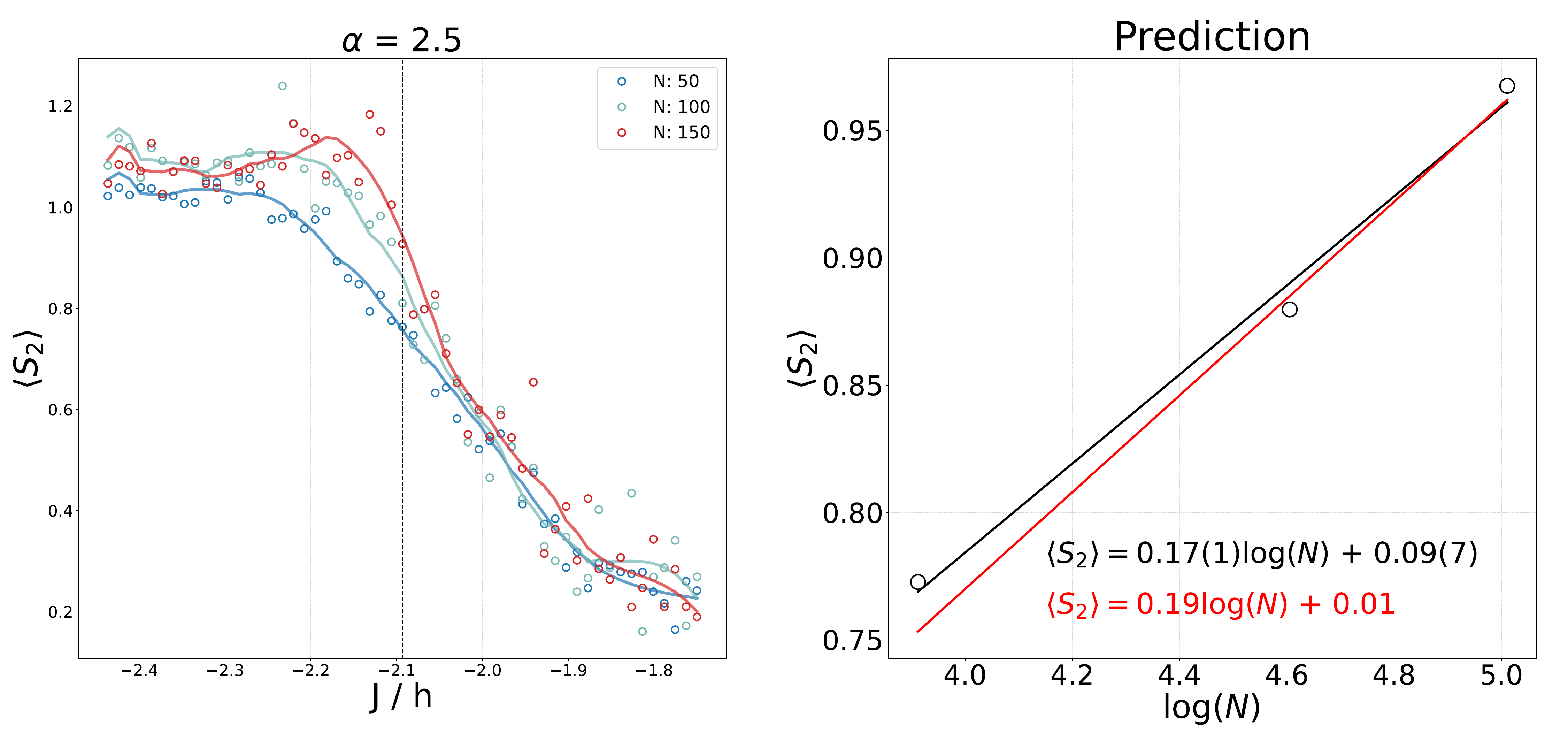}
    \caption{\textbf{Left:} dependence of the Rényi $2$-entropy on the $J$ parameter for three different chain sizes and a common value of $\alpha=2.5$. The vertical dotted line represents the critical point of the system, which marks the transition from a ferromagnetic ordering to a paramagnetic one. This model is trained on $\alpha\in\Bqty*{1.66, 6}$ with up to $1024$ samples per training point. Note that the results presented in this graph correspond to the predictions for a single data point, as opposed to the mean calculation over multiple realizations of each $\lbrace \alpha, J/h \rbrace$ pair. \textbf{Right:} variation of $\langle S_2 \rangle$ with the system size at the phase transition for $\alpha=2.5$. The black circles show the set-transformer predictions, the black line is the linear regression against these points and the red line is the results from \cite{PhysRevB.110.205147}}
    \label{fig:2.5entanglement}
\end{figure}

As explained in the previous section, for the second Rényi entanglement entropy, the set transformer was trained on a set of states where each pair $\lbrace \alpha, J/h\rbrace$ has five realizations, but each one of them is labeled by their mean entanglement entropy, instead of the individual values. This is done with the intention of averaging out the numerical fluctuations that arise from the QMC calculations, and at the same time enables us to get averaged outputs of a single state used as input  once the regressor is trained. Furthermore, we employ a transfer-learning technique, as explained in Section~\ref{sec:transferlearnign}, in order to reduce the training cost for the larger system sizes once the model has been trained on the shortest chain. The left panel of Figure~\ref{fig:2.5entanglement} shows the predictions of the set transformer for the states of $\alpha = 2.5$ for system sizes of $N = 50, 100, 150$, where we also used a Savitzky-Golay filter \cite{Savitzky1964}, which allows us to better capture the trend with increased size as we approach the critical point. To further understand the increase of the entanglement with the system size at the phase transition, we study their relationship. In the right panel of Figure~\ref{fig:2.5entanglement} we show the predicted logarithmic volume law calculated with the results of the set transformer. This result can be compared with those found in the literature, such as in \cite{PhysRevB.110.205147}, where the volume law reported was $\langle S_2 \rangle = 0.19\log(N)+0.01$.

To further assess the predictive power of the model, we analyze the finite-size scaling behavior of the magnetization fluctuations $\langle \hat{m}_s^2 \rangle$. Near the critical point, this quantity is expected to scale with the system size $N$ as
\begin{equation}
    \langle \hat{m}_s^2 (N, J) \rangle = N^{-2\beta/\nu} f\left(N^{1/\nu} (J - J_c)\right),
\end{equation}
where $\nu$ and $\beta$ are the standard critical exponents for the divergence of the correlation length and the order parameter, respectively. The function $f$ is a universal scaling function accounting for finite-size effects. To extract the values of $J_c$, $\nu$, and $\beta$, we perform a data collapse analysis using the Python library \software{fssa}~\cite{fssa, melchert2009autoscalepy}.

\begin{figure}[t]
    \centering
    \includegraphics[width=0.8\linewidth]{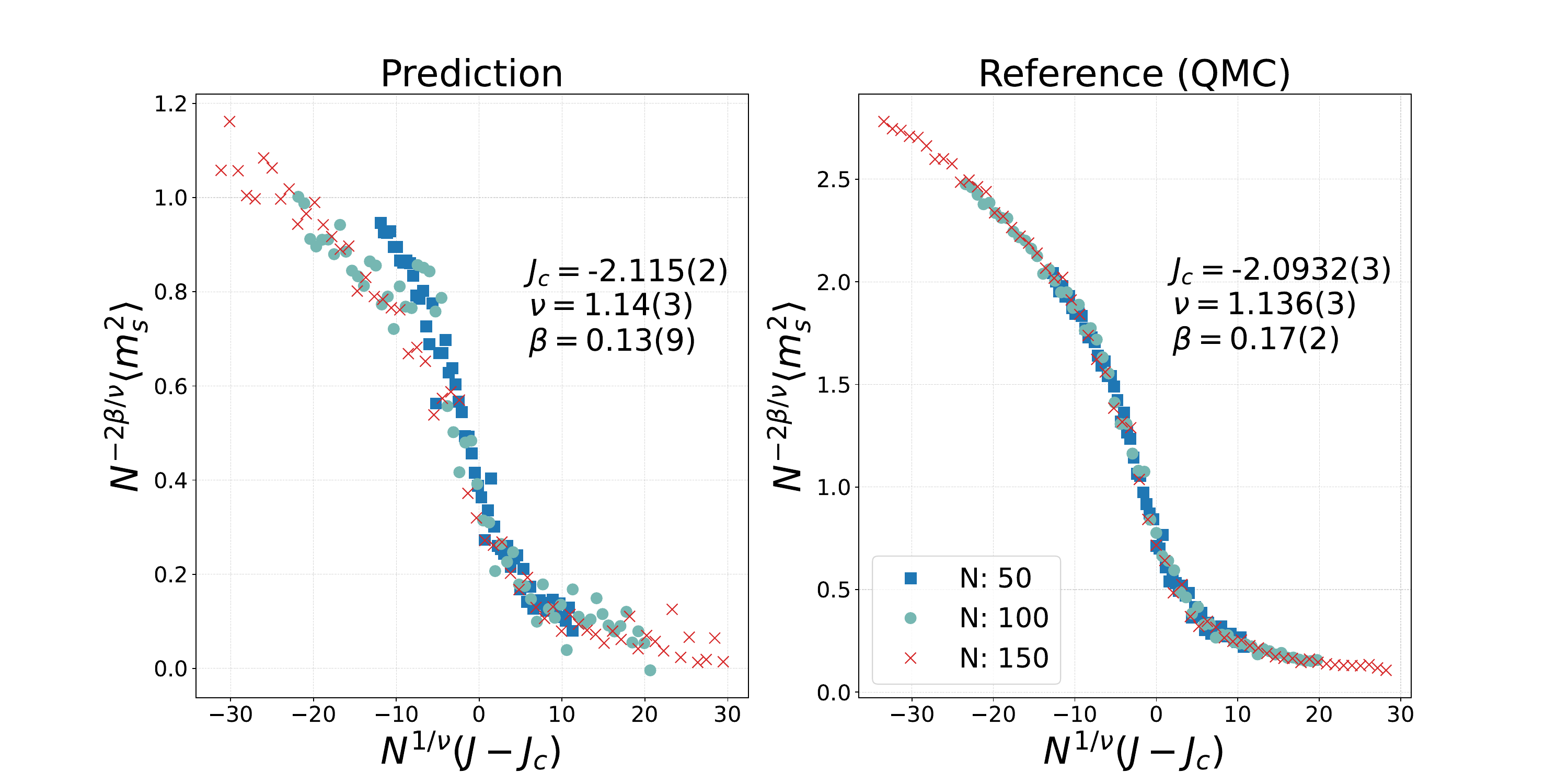}
    \caption{Collapses of $\aqty*{m_s^2}$ \textbf{Left:} collapse of the results predicted with the set transformer along with the predicted critical exponents $\beta, \nu$ and the critical point $J_c$. This model was trained on $\alpha\in\Bqty*{1.66, 6}$ with up to $1024$ samples per training point. \textbf{Right:} same set of results, calculated via QMC techniques.}
    \label{fig:collapse}
\end{figure}

Figure~\ref{fig:collapse} presents the data collapse obtained from the set-transformer predictions and from our QMC calculations, both based on the same sampled ground states. Although the visual collapse of the regressor results appears less sharp than the QMC counterpart, Figure~\ref{fig:collapse} shows that these results are consistent within the expected numerical uncertainties. Furthermore, the critical values extracted, summarized in Table~\ref{table:critical}, compare favorably with those reported in the literature.

\begin{table}[h!]
\centering
\resizebox{12cm}{!}{%
\begin{tabular}{cccc}
\toprule
\multicolumn{4}{c}{FM-PM Transition at $\alpha = 2.5$} \\
\midrule
Method & $J_c$  & $\nu$ & $\beta$ \\
\midrule
Set-Transformer (Ours) & $-2.115(2)$ & $1.14(3)$ & $0.13(9)$ \\
Our QMC Calculation     & $-2.0932(3)$ & $1.136(3)$ & $0.17(2)$ \\
Ref.~\cite{PhysRevB.110.205147} & $-2.09(1)$ & $1.08(9)$ & $0.19(3)$ \\
Ref.~\cite{koziol2021quantumcritical} & $-2.0878$ & $1.1084$ & $0.1880$ \\
\bottomrule
\end{tabular}%
}
\caption{Critical parameters extracted from finite-size scaling analysis of magnetization fluctuations.}\label{table:critical}
\end{table}

\section{Conclusion}

In this work, we have investigated the use of set-equivariant architectures ---specifically, set transformers--- for distinguishing quantum phases and predicting physical observables directly from raw quantum Monte Carlo samples. Our results show that this approach can efficiently extract relevant features from limited data, enabling accurate phase classification and the estimation of quantities such as magnetization moments and entanglement entropy, while accelerating the calculations by several orders of magnitude. Moreover, the proposed transfer learning workflow allows the model to be trained using only a fraction of the original dataset, thereby substantially reducing both the computational cost and the data acquisition time.

Furthermore, the model's predictions were sufficiently accurate to investigate the scaling behavior of the entanglement entropy with system size, yielding results consistent with previous studies. We have also performed a finite-size scaling analysis by collapsing the predicted values of the generalized magnetization, thereby extracting the critical coefficients and the critical point of the model. Although the data collapse is not perfectly sharp, the extracted values agree well with those reported in the literature.

The accuracy of the model and its predictions is inherently limited by the quality of the ground states used in the training set, as insufficiently converged quantum states introduce numerical noise. This noise may be inadvertently learned by the set transformer, effectively replacing relevant physical information with random patterns.

Our transfer-learning scheme facilitates a more efficient training process across different chain lengths, although it still relies on some amount for training data for each system size, and therefore does not enable the study of problem sizes beyond those represented in the dataset.

\ack

This study was supported by MCIN with funding from European Union NextGenerationEU (PRTR-C17.I1) promoted by the Government of Aragon. The authors acknowledge grant CEX2023-001286-S funded by MICIU/AEI /10.13039/501100011033.

\bibliographystyle{iopart-num-doi}
\bibliography{manuscript}
\end{document}